\def \SAIT #1 #2 {{\em Mem.\ Soc.\ Astron.\ It.\/} {\bf #1}, #2}
\def \MESS #1 #2 {{\em The Messenger\/} {\bf #1}, #2}
\def \ASTRNACH #1 #2 {{\em Astron. Nach.\/} {\bf #1}, #2}
\def \AAP #1 #2 {{\em Astron. Astrophys.\/} {\bf #1}, #2}
\def \AAL #1 #2 {{\em Astron. Astrophys. Lett.\/} {\bf #1}, L#2}
\def \AAR #1 #2 {{\em Astron. Astrophys. Rev.\/} {\bf #1}, #2}
\def \AAS #1 #2 {{\em Astron. Astrophys. Suppl. Ser.\/} {\bf #1}, #2}
\def \AJ #1 #2 {{\em Astron. J.\/} {\bf #1}, #2}
\def \ANNREV #1 #2 {{\em Ann. Rev. Astron. Astrophys.\/} {\bf #1}, #2}
\def \APJ #1 #2 {{\em Astrophys. J.\/} {\bf #1}, #2}
\def \APJL #1 #2 {{\em Astrophys. J. Lett.\/} {\bf #1}, L#2}
\def \APJS #1 #2 {{\em Astrophys. J. Suppl.\/} {\bf #1}, #2}
\def \APSS #1 #2 {{\em Astrophys. Space Sci.\/} {\bf #1}, #2}
\def \ASR #1 #2 {{\em Adv. Space Res.\/} {\bf #1}, #2}
\def \BAIC #1 #2 {{\em Bull. Astron. Inst. Czechosl.\/} {\bf #1}, #2}
\def \JSQRT #1 #2 {{\em J. Quant. Spectrosc. Radiat. Transfer\/} {\bf #1}, #2}
\def \MN #1 #2 {{\em Mon. Not. R. Astr. Soc.\/} {\bf #1}, #2}
\def \MEM #1 #2 {{\em Mem. R. Astr. Soc.\/} {\bf #1}, #2}
\def \PLR #1 #2 {{\em Phys. Lett. Rev.\/} {\bf #1}, #2}
\def \PASJ #1 #2 {{\em Publ. Astron. Soc. Japan\/} {\bf #1}, #2}
\def \PASP #1 #2 {{\em Publ. Astr. Soc. Pacific\/} {\bf #1}, #2}
\def \NAT #1 #2 {{\em Nature\/} {\bf #1}, #2}

\documentstyle[twoside]{memsait}
\input epsf.sty
\begin{opening}
\title{TIME DEPENDENT SPECTRA OF BLAZARS}
\author{Marco Chiaberge$^1$, Gabriele Ghisellini$^2$}
\institute{$^1$Dipartimento di Fisica Generale, V. Giuria, 1, Torino, Italy\\
$^2$Osservatorio di Brera, V. Bianchi, 46, Merate (Lecco), Italy}
\date{} 
\end{opening}

\begin{document}

\oddpagefooter{}{}{} 
\evenpagefooter{}{}{} 
\ 
\bigskip

\begin{abstract}
We calculate the time dependent electron distribution under the assumption
of continouos injection of new, relativistic particles, and assuming 
radiative cooling (by synchrotron and inverse Compton) and particle escape.
Resulting photon spectra are calculated taking into account the time 
delays introduced by the different light travel times across the source.
We apply these calculations to the varying X--ray spectrum of Mkn 421.
\end{abstract}

\section{Need of time dependent calculations}
Blazars vary violently at all wavelengths, with timescales as short as
hours--days.
This suggests that the injection mechanism and/or the cooling processes
operate on timescales shorter than the light crossing time $R/c$.
This implies simmetric light curves during rapid flares, as observed
during particularly intensive monitoring campaigns.
To interpret in detail the variability pattern at different wavelenghts
we need to study the time dependent behavior of the emitting particle 
distribution.

\section{The model: assumptions}
\begin{itemize}
\item The source, of typical dimension $R$, embedded by a tangled
magnetic field $B$, moves relativistically, and the radiation is beamed
with a Doppler factor $\delta$.
\item Relativistic electrons are injected ho\-mo\-ge\-neous\-ly throughout
the source for a time which can be less than $R/c$.
\item We consider Synchrotron and Self Compton cooling, and particle
escape. 
\item The electron distribution is found by solving the continuity equation:
$$
{\partial N(\gamma,t)\over \partial t} \, =\, 
{\partial \over \partial \gamma}\left[
\dot\gamma N(\gamma, t) \right] +Q(\gamma)-{ N(\gamma, t) \over t_{esc}}
$$
\end{itemize}
where $Q(\gamma)$ is the injection term, $\dot \gamma$ is the cooling
term, and $t_{esc}$ is the escape timescale of the particles,
assumed to be independent of their energy.
We solve this equation numerically, according to the scheme
proposed by Chang \& Cooper (1970).


Since the cooling and injection timescales are shorter than $R/c$, the
particle distribution evolves more rapidly than the light crossing time.
{\it The observer will see a convolution of different spectra, each 
produced in a different region of the source}.
Initially we see only the emission by fresh electrons located in the 
slice nearest to us. 
After a time $R/c$ we see the entire source: the back of it with fresh 
electrons, and the front of it with older electrons.


\section{Application to Mkn 421}

In May 1994 Mkn 421 underwent a X--ray flare during an high state of
the Tev emission (Macomb et al. 1995). 
Fig. 1 shows the overall spectra taken from Macomb et al. (1996)
fitted by two SSC spectra calculated with our program assuming
that the particle distribution reached equilibrium.
The two models differ only by the total injected power (factor 2)
and the $\gamma_{max}$ (factor 3).
Takahashi et al. (1996) studied the time lag
between hard and soft X--rays (ASCA data), finding a time delay of $\sim$1 hour.
The also note that the Tev flux varied with approximately the same
amplitude of the X--rays, while the optical flux remained quasi constant.
We qualitatively reproduce these features by assuming:
i) injection of a flat power electron distribution ($\propto \gamma^{-1.5}$) 
between $\gamma = 1000$ and $\gamma = 8\times 10^5$ for a time equal to $R/c$;
ii) $R=1.5\times 10^{16}$ cm, $B=0.07$ G, $\delta = 15.5$, 
$L_{inj}=0.6\times 10^{42}$ erg/s;
iii) the flaring emission corresponding to the fast varying electron 
distribution is summed to a constant component.
Fig. 2 shows the light curves at four different frequencies: note that
the soft X--rays lag the hard X--rays (approximately 1 hour), and the
optical flux remains quasi constant. The peak of the optical emission lags
the hard X--rays by $\sim$3 hours, while the Tev peak has a delay of 2 hours.
The seed photon for the TeV emission have IR frequencies,
and they lag the X--rays: this is the reason for the delay of the
TeV emission w.r.t. the X--rays.

\begin{figure}
\epsfxsize=14cm 
\epsfbox{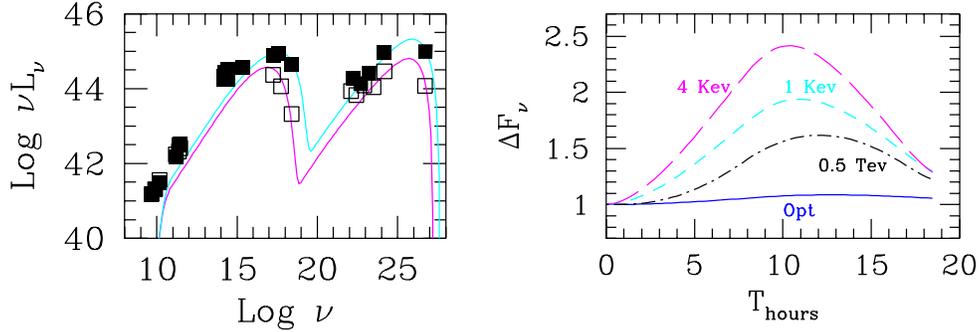} 
\vskip -9.75 true cm
\caption[h]{ {\it Left:} Overall spectra of Mkn 421 in the `quiescent' and
`flaring' state. Data from Macomb et al. (1996), fitted by
two SSC models. {\bf Fig. 2.}{\it Right:} 
Light curves at four different frequencies calculated for Mkn 421}
\end{figure}

\end{document}